\documentclass[twocolumn,aps,showpacs,prl,superscriptaddress]{revtex4} 

\usepackage{sidecap}
\usepackage{ulem}
\usepackage{epsfig}
\usepackage{amsmath,amssymb,amsthm}
\usepackage{graphicx}
\usepackage{bm}
\usepackage{color}

\DeclareMathAlphabet{\mathpzc}{OT1}{pzc}{m}{it}

\begin{document}

\renewcommand{\textfraction}{0.00}


\newcommand{\vAi}{{\cal A}_{i_1\cdots i_n}} 
\newcommand{\vAim}{{\cal A}_{i_1\cdots i_{n-1}}} 
\newcommand{\vAbi}{\bar{\cal A}^{i_1\cdots i_n}}
\newcommand{\vAbim}{\bar{\cal A}^{i_1\cdots i_{n-1}}}
\newcommand{\htS}{\hat{S}} 
\newcommand{\htR}{\hat{R}}
\newcommand{\htB}{\hat{B}} 
\newcommand{\htD}{\hat{D}}
\newcommand{\htV}{\hat{V}} 
\newcommand{\cT}{{\cal T}} 
\newcommand{\cM}{{\cal M}} 
\newcommand{\cMs}{{\cal M}^*}
\newcommand{\vk}{\vec{\mathbf{k}}}
\newcommand{\bk}{\bm{k}}
\newcommand{\kt}{\bm{k}_\perp}
\newcommand{\kp}{k_\perp}
\newcommand{\km}{k_\mathrm{max}}
\newcommand{\vl}{\vec{\mathbf{l}}}
\newcommand{\bl}{\bm{l}}
\newcommand{\bK}{\bm{K}} 
\newcommand{\bb}{\bm{b}} 
\newcommand{\qm}{q_\mathrm{max}}
\newcommand{\vp}{\vec{\mathbf{p}}}
\newcommand{\bp}{\bm{p}} 
\newcommand{\vq}{\vec{\mathbf{q}}}
\newcommand{\bq}{\bm{q}} 
\newcommand{\qt}{\bm{q}_\perp}
\newcommand{\qp}{q_\perp}
\newcommand{\bQ}{\bm{Q}}
\newcommand{\vx}{\vec{\mathbf{x}}}
\newcommand{\bx}{\bm{x}}
\newcommand{\tr}{{{\rm Tr\,}}} 
\newcommand{\bc}{\textcolor{blue}}

\newcommand{\beq}{\begin{equation}}
\newcommand{\eeq}[1]{\label{#1} \end{equation}} 
\newcommand{\ee}{\end{equation}}
\newcommand{\bea}{\begin{eqnarray}} 
\newcommand{\eea}{\end{eqnarray}}
\newcommand{\beqar}{\begin{eqnarray}} 
\newcommand{\eeqar}[1]{\label{#1}\end{eqnarray}}
 
\newcommand{\half}{{\textstyle\frac{1}{2}}} 
\newcommand{\ben}{\begin{enumerate}} 
\newcommand{\een}{\end{enumerate}}
\newcommand{\bit}{\begin{itemize}} 
\newcommand{\eit}{\end{itemize}}
\newcommand{\ec}{\end{center}}
\newcommand{\bra}[1]{\langle {#1}|}
\newcommand{\ket}[1]{|{#1}\rangle}
\newcommand{\norm}[2]{\langle{#1}|{#2}\rangle}
\newcommand{\brac}[3]{\langle{#1}|{#2}|{#3}\rangle} 
\newcommand{\hilb}{{\cal H}} 
\newcommand{\pleft}{\stackrel{\leftarrow}{\partial}}
\newcommand{\pright}{\stackrel{\rightarrow}{\partial}}

\title{Heavy flavor puzzle at RHIC: analysis of the underlying effects}

\author{Magdalena Djordjevic}
\affiliation{Institute of Physics Belgrade, University of Belgrade, Serbia}
\author{Marko Djordjevic}
\affiliation{Faculty of Biology, University of Belgrade, Serbia}

\begin{abstract} Suppressions of light and heavy flavor observables are considered to be excellent probes of QCD matter created in ultra-relativistic heavy ion collisions. Suppression predictions of quark and gluon jets appear to suggest a clear hierarchy according to which neutral pions should be more suppressed than D mesons, which in turn should be more suppressed than single electrons. However, joint comparison of neutral pion (light probe) and non-photonic single electron (heavy probe) suppression data at RHIC unexpectedly showed similar jet suppression for these two probes, which presents the well-known heavy flavor puzzle at RHIC. We here analyze which effects are responsible for this unexpected result, by using the dynamical energy loss formalism. We find that the main effect is a surprising reversal in the suppression hierarchy between neutral pions and D mesons, which is due to the deformation of the suppression patterns of light partons  by fragmentation functions. Furthermore, we find that, due to the decay functions, the single electron suppression approaches the D meson suppression. Consequently, we propose that these two effects, taken together, provide a clear intuitive explanation of this longstanding puzzle. 
\end{abstract}


\maketitle

\section{Introduction}
\begin{figure*}
\epsfig{file=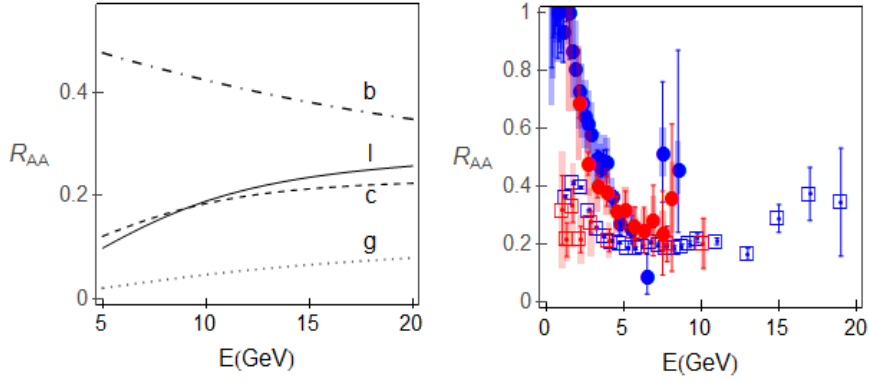,width=4.5in,height=2.2in,clip=5,angle=0}
\vspace*{-0.4cm}
\caption{{\bf The heavy flavor puzzle at RHIC.}  Momentum dependence of the jet suppression is shown on the left panel, for charm quarks (dashed curve), bottom quarks (dot-dashed curve), light quarks (full curve) and gluons (dotted curve). 
Electric to magnetic mass ratio is fixed to $\mu_M/\mu_E =0.4$, and the predictions are computed according to the Numerical framework section. The right panel shows together the experimentally measured 0-10\% central 200 GeV RHIC $R_{AA}$ data for neutral pions (open red squares from STAR~\cite{STAR_pi} and open blue squares from PHENIX~\cite{PHENIX_pi}) and non-photonic single electrons (full red circles from STAR~\cite{STAR_SE} and full blue circles from PHENIX~\cite{PHENIX_SE}).}
\label{PartonSuppRHIC}
\end{figure*}

\begin{figure*}
\epsfig{file=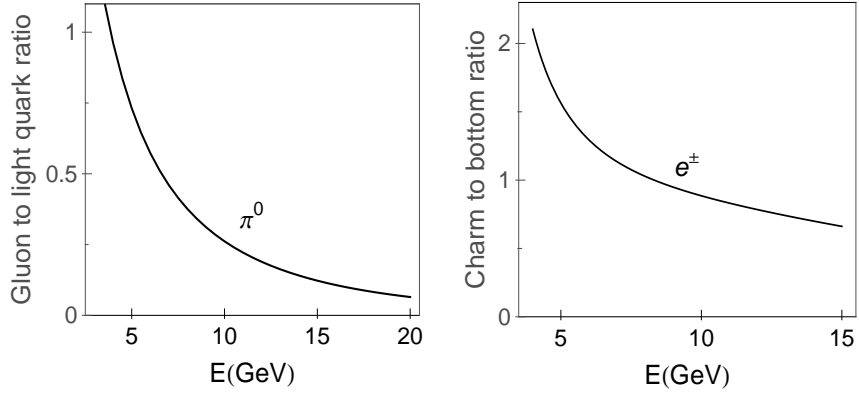,width=4.5in,height=2.2in,clip=5,angle=0}
\vspace*{-0.4cm}
\caption{{\bf Parton contribution in neutral pion and single electron production.} The left panel of the figure shows the gluon to light quark contribution ratio in the initial distributions of charged hadrons. The right panel of the figure shows charm to bottom quark ratio in the initial distributions of non-photonic single electrons. The ratio is computed according to the Numerical framework section.}
\label{PartonRatios}
\end{figure*}

Studying properties of QCD matter created in ultra-relativistic heavy ion collisions is a major goal of RHIC experiments. A powerful tool~\cite{Gyulassy,DBLecture,Wiedemann2013} to study these properties is suppression~\cite{Bjorken} of light and heavy flavor observables. It is intuitively expected that these observables should exhibit a clear hierarchy in the suppression patterns, which is based on the clear differences in the suppression of the underlying partons. The differences in the parton suppression can be clearly observed in the left panel of Fig.~\ref{PartonSuppRHIC}, which shows the suppression patterns for all types of quarks and gluons. From this figure, we see that charm and light quark suppressions are expected to be similar, but we also note that, due to steeper initial distributions of charm quarks, charm quark suppression is somewhat larger than light quark suppression, despite smaller charm quark energy loss. Furthermore, we see that, due to a larger color factor in the energy loss, pQCD predicts that gluon suppression should be significantly larger than for any other type of quarks, while due to a large mass (and consequently significant dead-cone effect~\cite{Kharzeev}), bottom quark suppression is significantly smaller than suppression for other partons. Furthermore,  Fig.~\ref{PartonRatios} shows that both light quarks and gluons significantly contribute to the neutral pion production, while both charm and bottom quarks significantly contribute to single electron production. The parton suppression in the left panel of Fig.~\ref{PartonSuppRHIC} then leads to the clear expectation for the probe suppression hierarchy: it is expected that pions should have a notably larger suppression than D mesons, which are, in turn, expected to have a significantly higher suppression than single electrons.

However, these intuitive expectations are clearly not supported by the measured data, which are shown in the right panel of Fig.\ref{PartonSuppRHIC}. This figure shows similar suppressions for neutral pions and single electrons, and this surprising result is called the heavy flavor puzzle at RHIC~\cite{HFPuzzle}. The puzzle has, up to now, inspired a significant amount of theoretical work~\cite{Gyulassy_Viewpoint} and even led to proposals that explaining the puzzle requires explanations outside of conventional pQCD\cite{FNG,NGT,GC,HGB}. 
The main goal of this paper is analyzing effects that are responsible for the heavy flavor puzzle at RHIC and consequently providing a clear intuitive explanation behind the puzzle.      

The analysis in this paper will be based on our dynamical energy loss formalism~~\cite{MD_PRC,DH_PRL,MD_Coll}, which was recently extended to finite magnetic mass~\cite{MD_MagnMass} and running coupling~\cite{MD_PLB}, and integrated in a numerical procedure for suppression predictions~\cite{MD_PLB}. Our approach in this analysis is based on the expectation that D meson suppression should be in-between pion and single electron suppressions. Consequently, to compare the pion suppression with single electron suppression - as relevant for the heavy flavor puzzle - we will first compare suppressions of neutral  pion and D meson, and then suppressions of D mesons and single electrons. We will show that the obtained (surprising) relative hierarchy can qualitatively explain the puzzle and consequently provide the desired intuitive explanation. Finally, we will also show that our most up-to-date numerical procedure can also provide an excellent quantitative explanation of the puzzling data.  

\section{Numerical framework} 
In our analysis, we will use our recently developed theoretical formalism, outlined in detail in~\cite{MD_PLB}. The procedure is based on {\it i)} radiative and collisional jet energy losses, computed in a finite size dynamical QCD medium~\cite{MD_PRC,DH_PRL,MD_Coll}, extended to the case of finite magnetic mass~\cite{MD_MagnMass} and running coupling~\cite{MD_PLB}, {\it ii)} multigluon~\cite{GLV_suppress} and path-length fluctuations~\cite{WHDG,Dainese} and {\it iii)} most up to date functions for  production~\cite{Vitev0912}, fragmentation~\cite{DSS} and decay~\cite{Cacciari:2012}.  

For RHIC conditions, we consider a QGP with 
$n_f{\,=\,}2.5$ effective light quark flavors and perturbative QCD scale of $\Lambda_{QCD}=0.2$~GeV. For the average temperature in our calculations, we use effective $T{\,=\,}221$\,MeV (as extracted by PHENIX~\cite{RHIC_T}). For charm and bottom masses we use, respectively, $M=1.2$~GeV and $M=4.75$~GeV.  For the light quarks, we assume  that their mass is dominated by the thermal mass $M{\,=\,}\mu_E/\sqrt{6}$, and the gluon mass is  $m_g=\mu_E/\sqrt{2}$~\cite{DG_TM}, where Debye mass $\mu_E \approx 0.7$~GeV is obtained by self-consistently solving the  Eq.~(3) from~\cite{MD_PLB} (see also~\cite{Peshier2006}). Magnetic mass $\mu_M$ is taken as $0.4 \, \mu_E < \mu_M < 0.6 \, \mu_E$~\cite{Maezawa,Bak}.  For all partons, the initial distributions are obtained from~\cite{Vitev0912}. For light hadrons, we use DSS fragmentation 
functions~\cite{DSS}. For D mesons we use BCFY fragmentation functions~\cite{BCFY}, while for B mesons we use KLP parameterization~\cite{KLP}. The decays of D, B mesons to non-photonic single electrons are obtained according to~\cite{Cacciari:2012}. Path length distributions are extracted from~\cite{Dainese}. Note that our computational procedure uses no free parameters, i.e. the parameters above correspond to the standard literature values. 

\section{Neutral pion {\it vs.} D meson suppression}

\begin{figure}
\epsfig{file=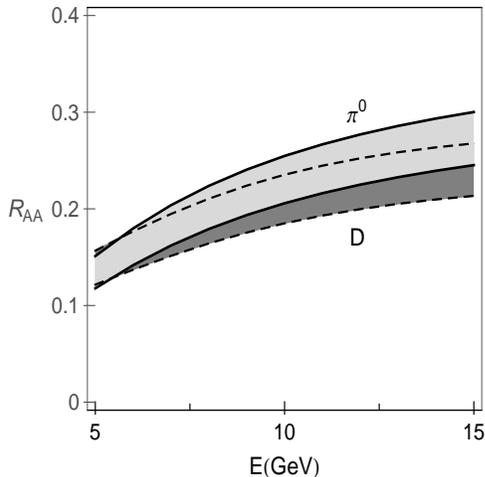,width=2.5in,height=2.5in,clip=5,angle=0}
\vspace*{-0.4cm}
\caption{{\bf Comparison of pion and D meson suppression predictions.} The figure  shows the comparison of neutral pion suppression predictions (light gray band)  with D meson (dark gray band) suppression predictions, as a function of momentum. Gray regions correspond to $0.4 < \mu_M/\mu_E < 0.6$, where the upper (lower) boundary on each band corresponds to $\mu_M/\mu_E =  0.6$ ($\mu_M/\mu_E =  0.4$).}
\label{PiDRaa}
\end{figure}

To understand the heavy flavor puzzle at RHIC, we first compare neutral pion with D meson suppression, as outlined in the Introduction. To this end, we use the dynamical energy loss formalism (see the previous section) to generate suppression predictions for neutral pions and D mesons. The predictions are shown in Fig.~\ref{PiDRaa}, where we directly compare the two suppressions. We see that, surprisingly, we obtain that D mesons should have a larger suppression compared to neutral pions. Note that this result is despite the fact that both light quarks and gluons significantly contribute to neutral pions (see the left panel of Fig.~\ref{PartonRatios} and that gluons have significantly higher suppression compared to both light quark and charm suppressions, while suppressions of light and charm quarks are similar. We will below analyze effects behind this unexpected result.

\begin{figure*}
\epsfig{file=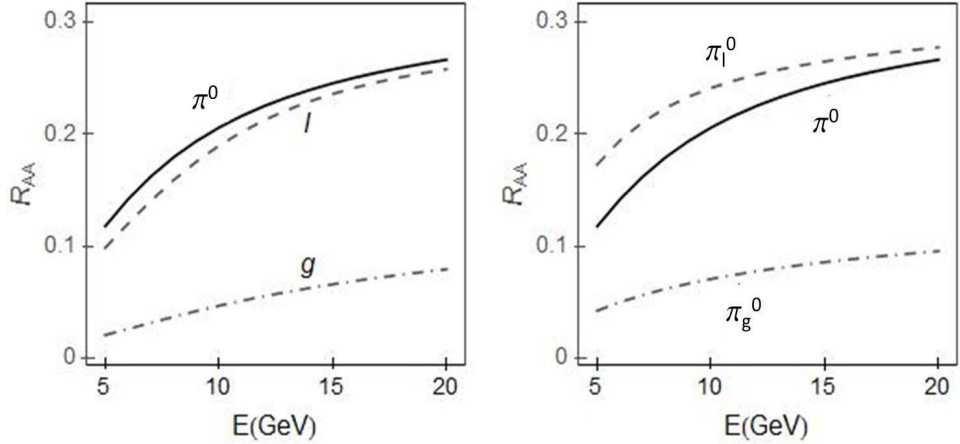,width=5in,height=2.5in,clip=5,angle=0}
\vspace*{-0.4cm}
\caption{{\bf Comparison of the light flavor suppression predictions.} The left panel  shows the comparison of neutral pion suppression predictions (full curve)  with light quark (the dashed curve) and gluon (the dot-dashed curve) suppression predictions, as a function of momentum. On the right panel, the dashed curve shows what would be the neutral pion suppression if only light quarks would contribute to pions. The dot-dashed curve shows what would be the neutral pion suppression if only gluons would contribute to pions, while the full curve shows the actual neutral pion suppression predictions. On each panel, electric to magnetic mass ratio is fixed to $\mu_M/\mu_E =0.4$.}
\label{PionQuarkRaa}
\end{figure*}

To this end, we next concentrate on how the fragmentation functions modify the parton suppressions, since these functions modify transfer from parton to hadron level. To study this, we first note that D       meson fragmentation  functions do not modify charm suppression, i.e. D meson suppression is  indeed a genuine probe of charm quark suppression~\cite{DGVW}. On the other hand, situation with neutral pions at RHIC is significantly more complicated: from the left panel of Fig.~\ref{PionQuarkRaa}, we see that pion fragmentation functions modify light quark and gluon suppressions in such a way that that their resultant neutral pion suppression is even smaller than bare light quark suppression. This counterintuitive result can be understood from the right panel in Fig.~\ref{PionQuarkRaa} and the left panel in Fig.~\ref{PartonRatios}. On the right panel of Fig.~\ref{PionQuarkRaa}, we plot what would be the suppression if pions were composed only of light quarks (dashed curve) and alternatively, what would be the suppression, if pions were composed only of gluons (dot-dashed curve). By comparing the left and the right panel in Fig. 4, we see that fragmentation functions significantly lower the suppression of its bare parton constituents (e.g. compare the dashed curves in these two figures). Furthermore, from the right panel of Fig.~\ref{PionQuarkRaa}, we see that pion suppression is much closer to the dashed curve (suppression if pions would consist only of light quarks), then to the dot-dashed curve (suppression if pions would consist only of gluons), which is due to the fact that the light quark contribution to pions dominates the gluon contribution. Consequently, lowering of the bare parton suppressions and dominance of the light quark contribution to pions, lead to (naively unexpected) smaller suppression of pions compared to light quarks, which is observed in the left panel of Fig.~\ref{PionQuarkRaa}. From this result, and the suppression hierarchy shown in Fig.~\ref{PartonSuppRHIC}, it follows that, at RHIC, high momentum D meson suppression should be larger than neutral pion suppression, as shown in Fig.~\ref{PiDRaa}; this result in itself presents an unintuitive reversal of expected hierarchy prediction to be tested against the upcoming high precision D meson $R_{AA}$ data from STAR.  

\section{Single electron {\it vs.} D meson suppression}

\begin{figure*}
\epsfig{file=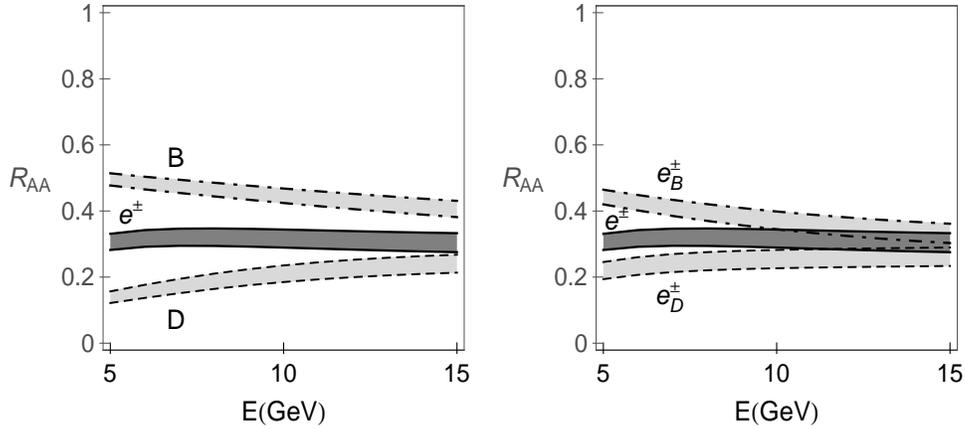,width=5in,height=2.5in,clip=5,angle=0}
\vspace*{-0.4cm}
\caption{{\bf Comparison of single electron with D and B meson suppression predictions.} The left panel shows the comparison of non-photonic single electron suppression predictions (dark-gray band with full curve boundaries) with D (light gray band with dashed curve boundaries) and B (light gray band with dot-dashed curve boundaries) meson suppression predictions, as a function of momentum.  The right panel shows the comparison of non-photonic single electron suppression predictions (dark-gray band with full curve boundaries) with single electron suppression from D mesons (light gray band with dashed curve boundaries) and single electron suppression from B mesons (light gray band with dot-dashed curve boundaries), as a function of momentum. Gray regions correspond to $0.4 < \mu_M/\mu_E < 0.6$, where the upper (lower) boundary on each band corresponds to $\mu_M/\mu_E =  0.6$ ($\mu_M/\mu_E =  0.4$).}
\label{DBSEPlot}
\end{figure*}

According to the outline in the Introduction, we next compare the single electron suppression with D meson suppression. While for single electrons (similarly as for D mesons) the fragmentation functions do not modify transfer from parton to hadron level~\cite{DGVW}, this transfer may be influenced by decay functions. To analyze this, we start by comparing our theoretical predictions for single electron $R_{AA}$ with $R_{AA}$s for D and B mesons, which is shown in the left panel of Fig.~\ref{DBSEPlot}.  
 Due to the fact that both D and B mesons significantly contribute to single electrons (see the right panel in Fig.~\ref{PartonRatios}), we see that the resultant single electron suppression (coming from the decay of these two mesons) is clearly in between these two suppression observables. However, we also note that single electron suppression is closer to D than to B meson suppression, despite the fact that, for higher momenta, B mesons dominate the single electron production (see the right panel of Fig.~\ref{PartonRatios}). To understand this, on the right panel, we plot what would be the single electron suppression if single electrons were composed only of D mesons (dashed band), and alternatively, what would be the single electron suppression if single electrons were composed only of B mesons (dot-dashed band). We see that actual single electron suppression is closer to the single electrons from B mesons, in agreement with the production ratio shown in the right panel of Fig.~\ref{PartonRatios}. However, by comparing these two panels, we also see that the decay functions modify D and B meson suppressions in such a way that their resultant single electron suppression is closer to D meson suppression. 

\section{Heavy flavor puzzle at RHIC}

\begin{figure*}
\epsfig{file=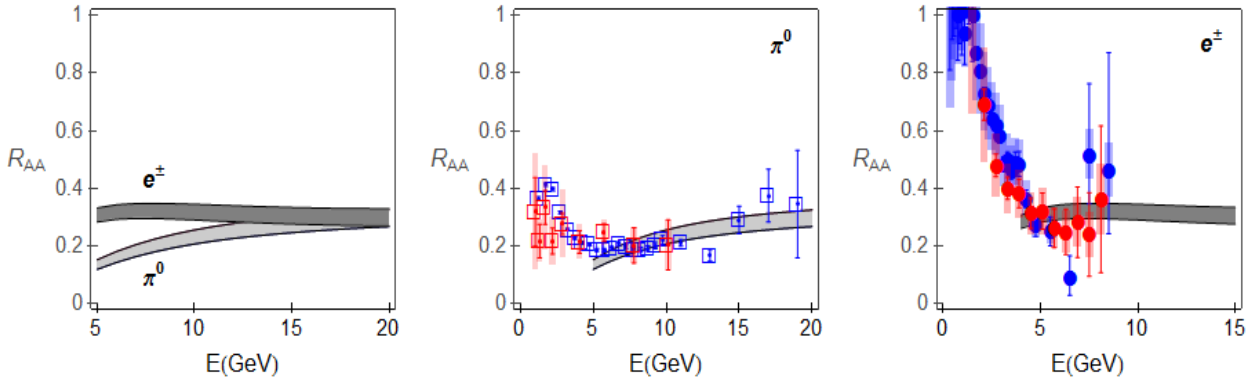,width=6.5in,height=2.2in,clip=5,angle=0}
\vspace*{-0.4cm}
\caption{{\bf Comparison of neutral pion and single electron suppression predictions with experimental data.}  The left  panel shows the comparison of neutral pion suppression predictions (light gray band) with non-photonic single electron suppression predictions (dark gray band).   The central  panel compares theoretical predictions for neutral pions (light gray band) with available pion $R_{AA}$ data at 0-10\% central 200 GeV RHIC (open red squares from STAR~\cite{STAR_pi} and open blue squares from PHENIX~\cite{PHENIX_pi}). The right panel compares theoretical predictions for single electrons (dark gray band) with the available RHIC single electron $R_{AA}$ data (full red circles from STAR~\cite{STAR_SE} and full blue circles from PHENIX~\cite{PHENIX_SE}). Gray regions correspond to $0.4 < \mu_M/\mu_E < 0.6$, where the upper (lower) boundary on each band corresponds to $\mu_M/\mu_E =  0.6$ ($\mu_M/\mu_E =  0.4$).}
\label{TheoryVsData}
\end{figure*}

In the analysis above, we obtained two important results, which directly lead to intuitive explanation of the heavy flavor puzzle at RHIC: We unexpectedly predicted that neutral pion suppression should be smaller than D meson suppression (Fig.~\ref{PiDRaa}) and that single electron suppression approaches D meson suppression (Fig.~\ref{DBSEPlot}). Taken together, these two results clearly lead to an expectation that single electron and neutral pion suppression patterns should approach each other. This expectation is confirmed in the left panel of Fig.~\ref{TheoryVsData}, where we show together the calculated single electron and neutral pion suppressions. While the left panel in Fig.~\ref{TheoryVsData} shows that our predictions qualitatively agree with the experimental data that form the heavy flavor puzzle at RHIC (see the right panel of Fig.~\ref{PartonSuppRHIC}), in the central and right panels of Fig.~\ref{TheoryVsData} we show a direct comparison of our theoretical predictions with, respectively, pion and single electron suppressions. Therefore, from Fig.~\ref{TheoryVsData}, we observe that we achieved both qualitative and quantitative agreement with RHIC pion and single electron suppression data, where we note that we use no free parameters in the model testing. 

\section{Conclusions}

A major theoretical goal of this paper was to analyze the effects that are responsible for the heavy flavor puzzle at RHIC, i.e. for the surprising experimental observation that the single electron suppression approaches the neutral pion suppression. This analysis is inherently quantitative, i.e. it involves interplay of energy loss, fragmentation and decay patterns. However, by comparing the suppression of pions and single electrons with that of D mesons, we found that the puzzle can be intuitively explained in terms of the following: {\it i)} we surprisingly predict that pion suppression should be smaller than D meson suppression, which is due to deformation of bare light quark and gluon suppressions by fragmentation functions, {\it ii)} we also found that, due to the deformation od D and B meson suppression patterns by the decay functions, single electron suppression approaches D meson suppression; this then inevitably leads to single electron suppression approaching the pion suppression, given the previous result that D meson suppression exceeds the pion suppression. This qualitative explanation is further complemented by a very good quantitative agreement of our model with measured neutral pion and single electron suppression data. Consequently, we argue that we provided both qualitative and quantitative understanding of the relevant data. We therefore conclude that pQCD description of the medium, and the corresponding calculations, can fully account for the heavy flavor puzzle at RHIC.

{\em Acknowledgments:} 
This work is supported by Marie Curie International Reintegration Grant 
within the $7^{th}$ European Community Framework Programme 
(PIRG08-GA-2010-276913) and by the Ministry of Science and Technological 
Development of the Republic of Serbia, under projects No. ON171004 and 
ON173052. We thank I. Vitev and Z. Kang for providing initial parton distributions and useful discussions.

\end{document}